\documentclass[twocolumn,showpacs,preprintnumbers,nofootinbib,nobibnotes,amsmath,amssymb,aps,prl]{revtex4-1}
\usepackage{todonotes}
\usepackage{graphicx}
\usepackage{hyperref}
\usepackage{slashed}
\usepackage{amssymb,graphicx}
\usepackage{epstopdf}
\usepackage{amsmath,amsfonts}
\usepackage{epsfig} 
\usepackage{graphicx,graphics}
\usepackage{xcolor}

\begin{document} 

\sloppy
\title{NANOGrav spectral index $\gamma=3$ from melting domain walls}

 \author{E.~Babichev$^{a}$, D.~Gorbunov$^{b,c}$, S. Ramazanov$^{d}$, R. Samanta$^{d}$, A.~Vikman$^{d}$\\
 \small{\em ${}^{a}$Universit\'e Paris-Saclay, CNRS/IN2P3, IJCLab, 91405 Orsay, France}\\
  \small{\em ${}^{b}$Institute for Nuclear Research of the Russian Academy of Sciences, 117312 Moscow, Russia}\\
    \small{\em ${}^{c}$Moscow Institute of Physics and Technology, 141700 Dolgoprudny, Russia}\\
\small{\em ${}^{d}$CEICO, FZU-Institute of Physics of the Czech Academy of Sciences,}\\
 \small{\em Na Slovance 1999/2, 182 00 Prague 8, Czech Republic}}

\begin{abstract}
We discuss cosmic domain walls described by a tension red-shifting with the expansion of the Universe. These melting domain walls emit gravitational waves with the low-frequency spectral shape $\Omega_{gw}\propto f^{2}$ corresponding to the spectral index $\gamma=3$ favoured by the recent NANOGrav 15 yrs data. We discuss a concrete high-energy physics scenario leading to such a melting domain wall network in the early Universe. This scenario involves a feebly coupled scalar field, which can serve as a promising dark matter candidate. We identify parameters of the model matching the gravitational wave characteristics observed in the NANOGrav data. The dark matter mass is pushed to the ultra-light range below $10^{-11}-10^{-12}\,\text{eV}$ which is accessible through planned observations thanks to the effects of superradiance of rotating black holes.

\end{abstract}

{\let\newpage\relax\maketitle}


{\it Introduction and Summary.} Recently several pulsar timing arrays (PTAs) such as NANOGrav \cite{ngr1,ngr2}, EPTA (including InPTA) \cite{epta1,epta2,epta3}, PPTA \cite{ppta1,ppta2}, and CPTA \cite{cpta} reported evidence for a common-spectrum signal in each dataset, with inter-pulsar angular correlations described by the Hellings-Downs (HD) curve \cite{hd}, pointing to a breakthrough discovery of nHz stochastic gravitational wave (GW) background. Signals from all the PTAs are in a good agreement, and in this article we shall focus on the NANOGrav 15yrs data with the largest statistical significance. The source of nHz GWs remains unknown, but the best fit power law GW spectrum $\Omega_{gw}\propto f^{1.2-2.4}$ at 68\% CL~\cite{ngr4} reported by NANOGrav\footnote{It is important to note that in 2020,  the NANOGrav reported similar common spectrum process in their 12.5 yrs dataset but without any evidence of HD correlation \cite{NANOGrav:2020bcs}. Compared to the old data, which are better fitted with a nearly scale-invariant spectrum: $\Omega_{gw}\propto f^{-1.5-0.5}$ at $1\sigma$, the 15 yrs data predict a much steeper spectrum, ruling out stable cosmic strings--one of the most anticipated primordial GW sources for PTAs \cite{cs1,cs2,cs3}.} disfavors simple GW-driven models of supermassive  black hole binaries (SMBHBs) predicting $\Omega_{gw}\propto f^{2/3}$ at 95\% CL \cite{ngr4, ngr3, Phinney:2001di}. While statistical and environmental effects may alleviate the tension \cite{ngr3,ngr4,bhn1,bhn2}, the latter motivates to investigate a possible cosmological origin of the observed GW signal. We explore this possibility in the present Letter and focus on GWs from cosmic domain walls (DWs)~\cite{Zeldovich:1974uw}.

 Compared to previous works, which fit constant tension DWs to the NANOGrav 15 yrs signal~\cite{Madge:2023cak,dom1,dom2,dom3, Sakharov:2021dim, dom4,dom5,dom6,dom7,dom8,dom9,dom10, dom11}, we consider so-called melting DWs characterized by a time-dependent tension, which drops as a cube of the Universe temperature~\cite{Vilenkin:1981zs, Babichev:2021uvl, Ramazanov:2021eya}. Such DWs are cosmology friendly, as their energy density redshifts fast enough not to overclose the Universe. They naturally arise in a well-motivated renormalizable particle physics scenario involving feebly coupled scalar field~\cite{Babichev:2021uvl, Ramazanov:2021eya}, which we briefly review in this Letter.
 
These {\it melting} DWs 
serve as a source of GWs with the spectrum distinguishable from the one provided by constant tension DWs. Most notably, the low-frequency GW spectrum from melting DWs behaves as\footnote{Among other topological defects, the network of metastable cosmic strings is also capable of producing the low-frequency GW spectrum $\Omega_{gw} \propto f^2$~\cite{Buchmuller:2023aus}.} $\Omega_{gw} \propto f^2$~\cite{Babichev:2021uvl}, which should be compared with $\Omega_{gw} \propto f^3$~\cite{Hiramatsu:2013qaa} in the case of constant tension walls. 
The larger signal at small frequencies stems from the fact that the network of melting DWs efficiently emits GWs over an extended period of time: while the most energetic GWs are produced at the network formation, later emission from somewhat melted DWs feeds into the low energy tail of the spectrum. This contrasts sharply with the constant tension case, where GWs are mainly emitted at the end of wall evolution right before dissolving, e.g., due to slight breaking of $Z_2$-symmetry. Note that there is no contradiction with causality considerations~\cite{Caprini:2009fx, Cai:2019cdl, Hook:2020phx, Durrer:2003ja, Franciolini:2023wjm}, which typically lead to $\Omega_{gw} \propto f^3$. Indeed, the standard steeper shape assumes a finite operation of the GW source, shorter than the Hubble time. In contrast, in the scenario~\cite{Babichev:2021uvl, Ramazanov:2021eya} we discuss here, GWs are efficiently produced by the time-extended source over many Hubble time intervals.

Remarkably, the behaviour $\Omega_{gw} \propto f^2$ better fits NANOGrav 15 yrs data compared to\footnote{At the same time, low-frequency GW emission from melting cosmic strings has a shape $\Omega_{gw} \propto f^4$~\cite{Emond:2021vts}, which conflicts with NANOGrav data.} $\Omega_{gw} \propto f^3$. It is conventional to parameterise the PTA GW signal as 
\begin{equation}
\label{ng3}
\Omega_{gw}(f)=\Omega_{yr}\left(\frac{f}{f_{yr}}\right)^{5-\gamma}, 
\end{equation}
with $\gamma$ being the spectral index and $f_{yr}={\rm 1~yr^{-1}}\simeq 32$ nHz and $\Omega_{yr}=5.8\times 10^{-8}$. The NANOGrav best-fit value of the spectral index reads $\gamma =3.2\pm 0.6$ \cite{ngr1}, which is automatically recovered in the scenario with melting DWs. We demonstrate that matching to other characteristics of the NANOGrav 15 yr signal, i.e., maximal frequency $f_{yr}$ and energy density $\Omega_{yr}$, allows us to unambiguously define coupling constants of the scalar field constituting melting DWs. In particular, this scalar field should be extremely 
weakly coupled making it a suitable dark matter candidate, provided that its mass is confined to the ultra-light range. For such low masses, superradiance~\cite{zeldovich1,zeldovich2,Starobinsky:1973aij} plays an important role by triggering instability of rotating black holes with astrophysical masses~\cite{Arvanitaki:2009fg}. 
This leads to potentially observable spin-down of rotating black holes and to stochastic GW background due to gravitational radiation of the bosonic condensate forming around black holes, see, e.g., Ref.~\cite{Brito:2015oca}.


{\it Overview of melting domain walls scenario.} The scenario giving rise to melting DWs involves the $Z_2$-symmetric model of a real scalar field $\chi$, 
which interacts through the portal coupling with a scalar multiplet $\phi$ from the primordial thermal bath:
\begin{equation}
\label{model}
{\cal L}=\frac{(\partial_{\mu} \chi)^2}{2}  -\frac{M^2_{\chi} \chi^2}{2}-\frac{\lambda_{\chi} \chi^4}{4} +\frac{g^2 \chi^2 |\phi|^2}{2}  \; ,
\end{equation}
where $M_{\chi}$, $\lambda_{\chi}$, and $g^2$ are the bare mass, quartic self-interaction constant of the field $\chi$, and portal coupling constant, respectively~\cite{Babichev:2021uvl, Ramazanov:2021eya}. We assume that particles $\phi$ are relativistic at the times of interest, which fixes the variance of the field $\phi$ to be $\langle |\phi|^2 \rangle =\frac{N T^2}{12}$, where $N$ counts the number of degrees of freedom associated with $\phi$. The sign of the portal coupling in the scenario is fixed as $g^2>0$. 
This induces tachyon instability in the two-field system, which is tamed, provided that the following condition is obeyed: 
\begin{equation}
\label{beta}
\beta \equiv \frac{\lambda_{\chi}}{g^4} \geq \frac{1}{\lambda_{\phi}} \; ,
\end{equation}
where $\lambda_{\phi}$ is the quartic self-interaction constant of the multiplet $\phi$ (The ratio of the coupling constants $\beta$ we introduced above will appear in the relations below). As a result, the effective potential characterizing the field $\chi$ exhibits spontaneous symmetry breaking leading to the non-zero temperature-dependent expectation value: 
\begin{equation}
\label{expect}
\langle \chi \rangle =\pm \sqrt{\frac{Ng^2 T^2}{12\lambda_{\chi}} -\frac{M^2_{\chi}}{\lambda_{\chi}}} \; .
\end{equation}
In the expanding Universe, this temperature-dependence induces time-dependence, which is crucial for our further discussions. At some (lower) temperature the bare mass term becomes relevant, and the symmetry restores with $\langle \chi \rangle=0$, i.e., the inverse phase transition happens. However, at the times of the cosmological evolution we are interested, the bare mass $M_{\chi}$ is negligible; it will be included only when considering dark matter implications of the model. 

Spontaneous breaking of $Z_2$-symmetry leads to the formation of DWs, provided that the background field $\chi$ is set to zero, i.e., $\chi=0$, prior to falling into the minima of symmetry breaking potential; see Ref.~\cite{Babichev:2021uvl} for details. DWs are often unwelcome in cosmology because they quickly begin to dominate the evolution of the Universe, in contradiction with observational data. This problem is absent in our case, exactly due to the time dependence of the expectation value $\langle \chi \rangle $, as it will become clear shortly. 

The Universe temperature at the time of DW formation is defined by the balance of the Hubble friction and 
the tachyonic thermal mass; it is estimated as 
\begin{equation}
\label{Ti}
T_i \simeq \frac{\sqrt{N} gM_{Pl}}{\sqrt{B g_* (T_i)}} \; ,
\end{equation}
where $g_* (T)$ counts the number of relativistic degrees of freedom at the temperature $T$, and $M_{Pl} \approx 2.44 \cdot 10^{18}~\mbox{GeV}$ is the reduced Planck mass. The constant $B$ here takes into account the finite duration of the field $\chi$ roll to the minimum of its potential; $B \simeq 1$ for the infinitely fast roll, 
but generically it takes values in the range $1 \lesssim B \lesssim 10^3$, see Ref.~\cite{Babichev:2021uvl}. 

The DW tension (mass per unit area) is given by $\sigma =2\sqrt{2 \lambda_{\chi}} \cdot \langle \chi \rangle^3/3$. The energy density of DWs in the scaling regime with one (a few) DWs per "horizon" volume $H^{-3}$, 
where $H$ is the Hubble rate, is estimated as $\rho_{wall} \sim \sigma H$. Using Eq.~\eqref{expect}, where we neglect the bare mass, one finds that the energy density of melting DWs redshifts as $\rho_{wall} \propto 1/a^5$ at the radiation-dominated stage, which is in contrast to the scenario with constant tension DWs yielding $\rho_{wall} \propto 1/a^2$. Hence, the energy density of melting DWs drops faster  than the energy density of radiation, and there is no DW problem in the Universe.

{\it GWs from melting domain walls.} In both scenarios with constant tension and melting DWs, most energetic GWs are emitted within a short time interval (of the order of the Hubble time): close to the moment of the DW formation $t_i$~\cite{Babichev:2021uvl} in the case of melting DWs and near the time of 
the network dissolution for constant tension DWs~\cite{Hiramatsu:2013qaa}. In this time interval, which is crucial for defining peak properties of GWs, one can approximate the tension of melting DWs as constant, and use some of results of analysis of Ref.~\cite{Hiramatsu:2013qaa} supported by numerical simulations performed assuming constant tension DWs. In particular, the peak frequency is estimated by the Hubble rate at the time $t_i$, i.e., $H_i$. Taking into account the redshift, the present-day peak frequency reads~\cite{Babichev:2021uvl, Ramazanov:2021eya}
\begin{equation}
\label{peakfr}
f_{p} \equiv f_{p} (t_0)  \simeq H_i  \cdot \frac{a_i}{a_0} \propto T_i \; ,
\end{equation}
which gives upon substituting of Eq.~\eqref{Ti}:
\begin{equation}
\label{peakfrequency}
f_{p} \simeq 6~\mbox{nHz}~\sqrt{\frac{N}{B}} \cdot \frac{g}{10^{-18}} \cdot \left(\frac{100}{g_* (T_i)} \right)^{1/3} \; .
\end{equation}
Similarly, a simple estimate based on dimensional analysis gives a rather accurate prediction for the fractional spectral energy density of GWs at peak~\cite{Hiramatsu:2013qaa}. Including correction factors taken from simulations of Ref.~\cite{Hiramatsu:2013qaa}, one can write
\begin{equation}
\label{peakenergyti}
\Omega_{gw, p} (t_i) \approx \frac{\lambda_{\chi} \epsilon_{gw} {\cal A}^2 \langle \chi \rangle^6_i}{27\pi H^2_i M^4_{Pl}} \propto T^2_i \; ,
\end{equation}
where the coefficients $\epsilon_{gw}$ and ${\cal A}$ account for the efficiency 
of GW emission and scaling, correspondingly; one has $\epsilon_{gw} {\cal A}^2 \approx 0.5$. In essence, this value of $\epsilon_{gw} {\cal A}^2$ is the only input from lattice simulations with constant tension walls, while the rest is fixed on dimensional grounds. Unless simulations with melting DWs show a considerably lower value of $\epsilon_{gw} {\cal A}^2$, our further discussion is unaffected. Combining Eqs.~\eqref{expect},~\eqref{Ti},~\eqref{peakenergyti}, using definition~\eqref{beta}, and taking into account the redshift of GW fractional energy density during matter-dominated stage, we obtain~\cite{Babichev:2021uvl, Ramazanov:2021eya}
\begin{equation}
\label{peakenergy}
\Omega_{gw, p}\, h^2_0 \simeq \frac{4 \cdot 10^{-14} \cdot N^4}{B\cdot \beta^2} 
\cdot \left(\frac{100}{g_* (T_i)} \right)^{7/3} \; ,
\end{equation}
where $h_0=0.67$ is the reduced Hubble constant \cite{Planck:2018vyg}. Note that the peak frequency and energy density of GWs are largely determined by the coupling constants of the field $\chi$; this property will be exploited later to recover these constants using PTA observations.

To discriminate between GWs emitted by melting domain walls and other sources, one should obtain the GW spectrum. For this purpose we observe that GW emission 
coming from the later times $t>t_i$ occurs at the characteristic frequency $f \simeq H(t) a(t)/a_0 \propto T(t)$, --- in full analogy with Eq.~\eqref{peakfr}; 
it is obvious that $f<f_p$.
Furthermore, using the same considerations, which led to Eq.~\eqref{peakenergyti}, one obtains that the fractional energy density of GWs emitted at the times $t$ behaves as $\Omega_{gw} (t) \propto T^2 (t)$. Assuming that these GWs sourced at the times $t$ give the main contribution to the spectral energy density at frequency $f$, we obtain the low frequency part of the spectrum~\cite{Babichev:2021uvl}:
\begin{equation}
\label{lowfr}
\Omega_{gw}  \left(f<f_{p}  \right) \simeq \Omega_{gw, p}  \cdot \left( \frac{T(t)}{T_i} \right)^2 \simeq  \Omega_{gw, p}  \cdot \left( \frac{f}{f_{p}} \right)^2 \; .
\end{equation}
For more details see Appendix, where we prove that GW emission coming from the times earlier and later than $\sim t$ leaves the behaviour in Eq.~\eqref{lowfr} intact\footnote{See also Ref.~\cite{Ramazanov:2023eau} for an alternative proof and generalization to other types of long-lasting GW sources.}. Note that Eq.~\eqref{lowfr} is in contrast to the result obtained in the case of 
constant tension DWs and many other sources, 
i.e., first-order phase transitions and (stable) cosmic strings, giving $\Omega_{gw}h^2_0\propto f^3$. This does not imply violation of causality: indeed, according 
to the discussion above, low-frequency GW emission still fulfills $f \simeq H(t)a(t)/a_0$ and hence follows from on-horizon dynamics of melting DWs. Concerning the high-frequency part of the spectrum, we do not study it in detail, since in our scenario it is outside of the domain probed by NANOGrav (see below). We may safely assume that it is not different from the case of constant tension walls, i.e., there is a power law decrease $\Omega_{gw} \propto 1/f$ at $f>f_{p}$ followed by the exponential suppression at frequencies corresponding to the inverse width of DWs~\cite{Hiramatsu:2013qaa}. 

\begin{figure*}
    \includegraphics[scale=0.39]{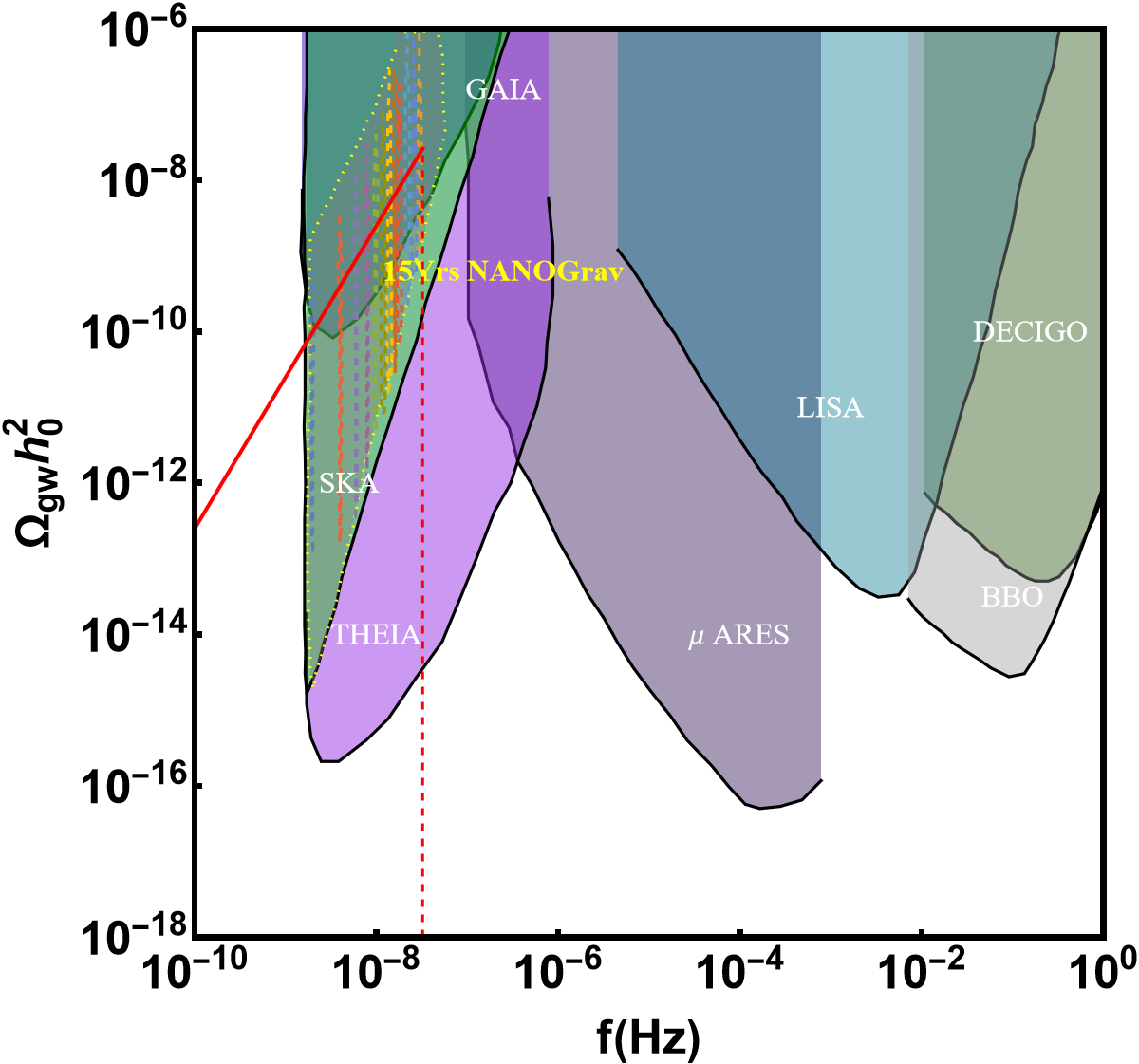}   \includegraphics[scale=0.38]{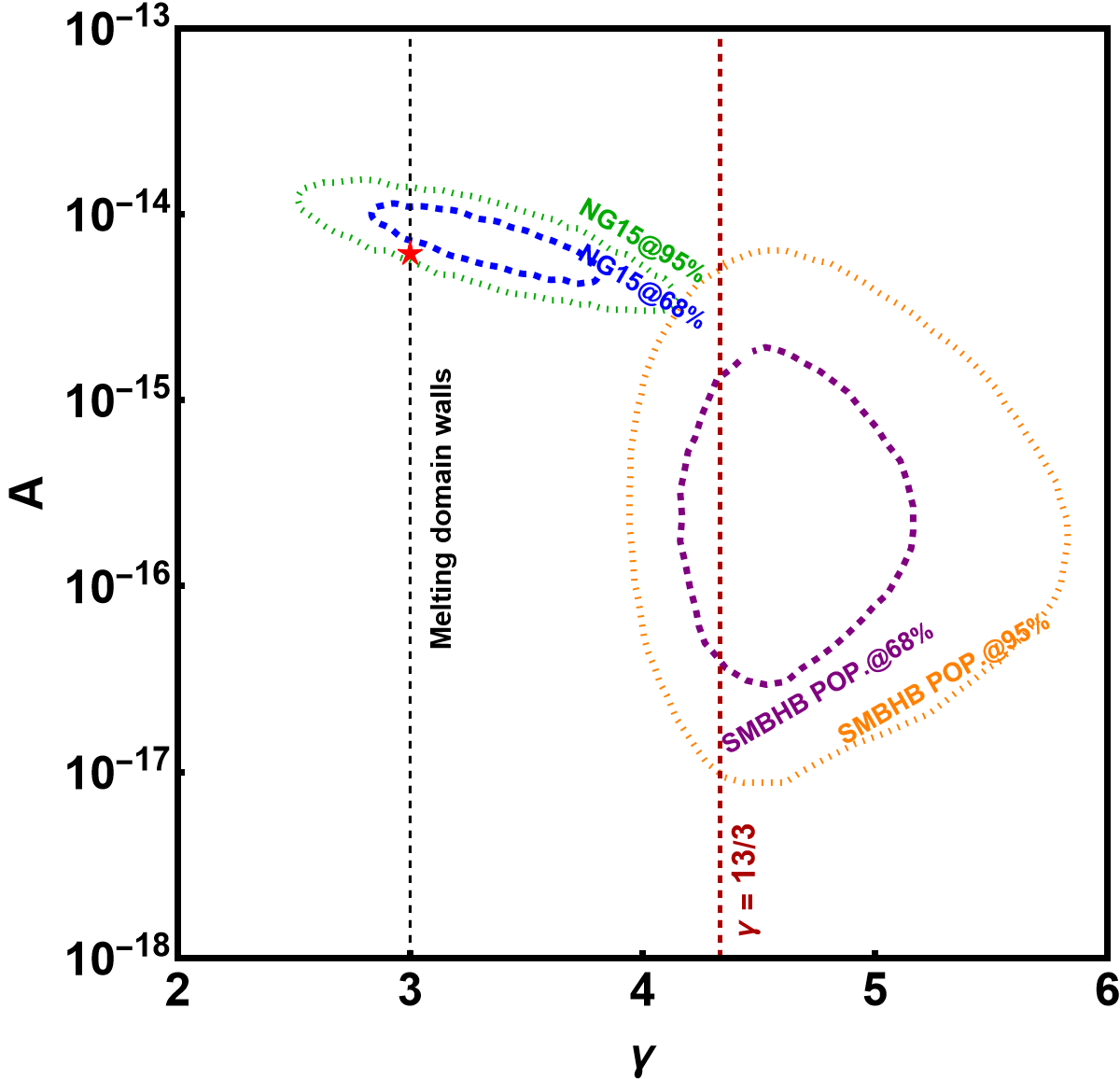}\\
     \includegraphics[scale=0.78]{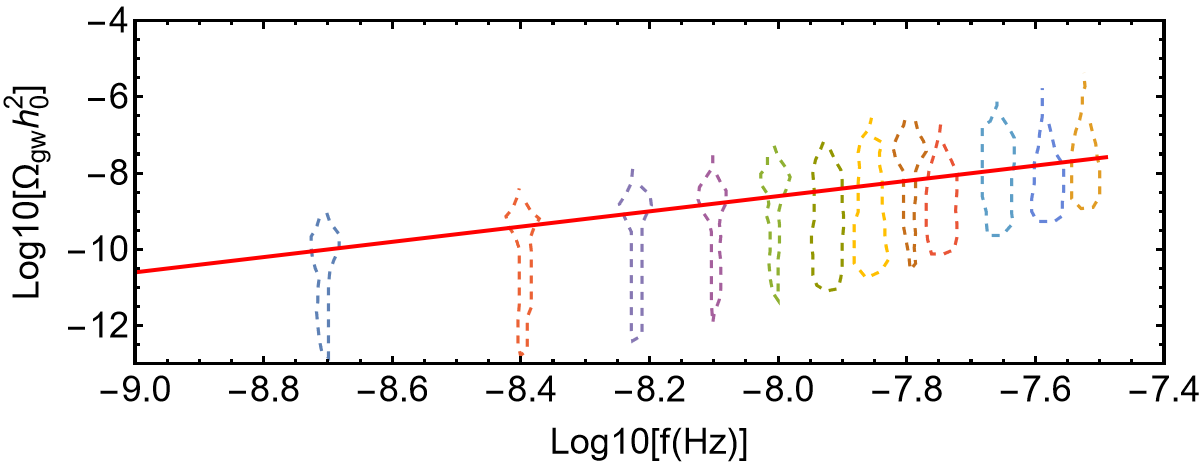}
    \caption{Top-Left: Spectral shape of GWs emitted by melting domain walls is shown with the solid red line versus sensitivity curves of various current and planned PTAs and GW interferometers. The plot has been produced for $g= 10^{-18}$, $\beta=1$, $B=1$, $N=24$ and $g_*(T_i)=75$. The spectrum is cut off at $f_{p} \simeq f_{yr}$ (vertical red dashed line). The shaded regions show the sensitivity of SKA \cite{ska}, GAIA and THEIA \cite{gaia}, $\mu$ARES \cite{mras}, LISA \cite{lisa}, DECIGO \cite{decigo}, and BBO \cite{bbo}. Top-Right:  $68\%$ and $95\%$ CL regions for the amplitude $A$ and spectral index $\gamma$ of a power-law fit to the observed GW signal (blue and green contours, respectively). At the reference frequency $f=f_{\rm yr}$, the NANOGrav best-fit values are $A\simeq 6.4^{+4.2}_{-2.7}\times 10^{-15}$ and $\gamma=3.2 \pm 0.6$.
    The amplitude and spectral index, $A\simeq 6.3\times 10^{-15}$ and $\gamma=3$, predicted for the same set of model parameters as in the left plot, are marked with the red `star.' For comparison, we show $68\%$ and $95\%$ CL regions for $A$ and $\gamma$ predicted for GW-driven supermassive black hole binary populations with circular orbits (purple and orange contours, respectively), and the best-fit value $\gamma=13/3$ with the red dashed line \cite{Phinney:2001di}. Bottom: Zoomed-in plot of the GW spectrum against the NANOGrav 15yrs data (dashed violins) \cite{ngr1} within the frequency range $f\in \left[10^{-9},f_{\rm yr}\right]$ Hz.}
    \label{fig:enter-label}
\end{figure*}

Figure~\ref{fig:enter-label} demonstrates that the predicted GW signal is compatible with the NANOGrav signal for the set of consistent values of model parameters. 
Below, we explain the notations used and the assumed choice of model constants. GW spectral energy density
associated with the NANOGrav, or more generally, with the PTA signal is conventionally parameterised by the amplitude $A$ related to $\Omega_{yr}$ in Eq.~\eqref{ng3} by 
\begin{equation}
A =\sqrt{\frac{3\Omega_{yr} H^2_0}{2\pi^2 f^2_{yr}}} \; ,
\end{equation}
where $H_0$ is the Hubble constant; recall that $f_{\rm yr}={\rm 1~yr^{-1}}\simeq 32$ nHz. The best fit to the NANOGrav signal is provided by the values $A\simeq 6.4^{+4.2}_{-2.7}\times 10^{-15}$ and $\gamma=3.2 \pm 0.6 $ \cite{ngr1}. The latter automatically agrees well with the model prediction~\eqref{lowfr}.

To achieve the best fit value of $A$, which corresponds to rather large GW energy density $\Omega_{yr} \simeq 5.8 \times 10^{-8}$, we first set $f_{p} \simeq f_{yr}$, so that $\Omega_{gw, p} \simeq \Omega_{yr}$. The reason for this choice will become clear a posteriori. Combining Eqs.~\eqref{Ti} and~\eqref{peakfrequency}, we obtain $T_i \simeq 1.3~\mbox{GeV}$ and $g_* (T_i) \simeq 75$. Now we fix the constants entering the GW energy density~\eqref{peakenergy}: $ \beta = 1\,,~B = 1\, ,~N=24$. Finally, using this and Eq.~\eqref{peakfrequency}, we can fix the constant $g$, i.e., $g=10^{-18}$. This implies a tiny portal coupling $g^2=10^{-36}$, 
while $\beta=1$ translates into the self-interaction constant $\lambda_{\chi}=10^{-72}$. This demonstrates the point made in Ref.~\cite{Ramazanov:2021eya} that GWs from melting DWs can be used to identify otherwise inaccessible extremely weak interactions. Such tiny coupling constants are not unfamiliar in physics; they are characteristic for axion-like particles \cite{Choi:2020rgn}. However, the two setups have different symmetry properties, and thus should not be confused. Note also 
that we chose the constants $\beta$ and $B$ to be 
close to the lower bounds of allowed values,
see Eq.~\eqref{beta}, in order to achieve the observed value $\Omega_{yr}$. This also explains the choice $f_{p} \simeq f_{yr}$, because for $f_{p} \gg f_{yr}$, one would need to assume too large $\Omega_{gw, p} \gg \Omega_{yr}$. It is important to stress that one can accommodate larger values of parameters $\beta$ and $B$ by a moderate increase of the number of degrees of freedom $N$. Indeed, increasing $\beta$ by factor $\xi$ requires only an increase of $N$ by smaller factor $\xi^{1/2}$. On the other hand, a change of parameter $B$ by factor $\zeta$ requires a  corresponding increase of $N$ by a much smaller factor $\zeta^{1/4}$.

Let us comment on the properties of the field $\phi$. Our interpretation of PTA signal in terms of GWs from melting DWs bounds the mass $m_{\phi}$ of the field $\phi$ as $m_{\phi} \ll 1~\mbox{GeV}$, which is enforced by the requirement that $\phi$ is relativistic at the times of DWs formation. As a result, one runs the risk of spoiling a well established picture of BBN. There are two ways of avoiding this. One is to assume that the 
particles $\phi$ decoupled from primordial plasma at very early times, 
and thus contribute insignificantly to the effective number of neutrino species $N_{eff}$. In that case, however, the effective temperature $T_{\phi}$ describing the system of particles $\phi$ is lower than the Universe temperature. This tends to decrease GW energy density according to Eq.~\eqref{peakenergyti}, but the decrease can be (partially) compensated by the sharp change of degrees of freedom number $g_* (T)$ around QCD phase transition. Another way to handle the problem is to assume that the particles $\phi$ have mass $m_{\phi}$ in the MeV range, i.e., $1~\mbox{MeV} \ll m_{\phi} \ll 1~\mbox{GeV}$. That is, the particles $\phi$ become non-relativistic sometime before BBN and then decay into SM species in one or another way. In that case, one can also consider the scenario with the effective temperature $T_{\phi}$ higher than the Universe temperature $T$.

{\it Implications for dark matter.} The field $\chi$ being very feebly coupled to the primordial thermal bath is a proper dark matter candidate. It should be stressed that for such a tiny portal coupling, $g^2 \simeq 10^{-36}$, neither freeze-out nor freeze-in production mechanisms are efficient. Yet it is possible to generate the right dark matter abundance even with this tiny coupling constant. We briefly comment on two production mechanisms below and identify the mass $M_{\chi}$ as a function of GW parameters assuming that the field $\chi$ constitutes all dark matter.

i) Dark matter production via the direct phase transition~\cite{Babichev:2021uvl}. Oscillations of the field $\chi$ around the minima of its potential naturally feed into dark matter. These oscillations start at the times $t \simeq t_i$, when the DW network is created, and continue till present unless the particles $\chi$ are unstable. In that case, the observed dark matter abundance is achieved for extremely small $M_{\chi}$: 
\begin{equation}
M_{\chi} \simeq  10^{-16}~\mbox{eV}  \left(\frac{f_{p}}{30~\mbox{nHz}} \right)   \cdot \left(\frac{g_* (T_i)}{100} \right)^{1/6} \sqrt{\frac{10^{-8}}{\Omega_{gw, p}}} \; .\label{M_x_direct}
\end{equation}

ii) Dark matter via inverse phase transition~\cite{Babichev:2021uvl}, cf. Refs.~ \cite{Babichev:2020xeg, Ramazanov:2020ajq}. Dark matter production 
occurs also in the case, when there is an efficient decay channel for the aforementioned oscillations, and the field $\chi$ settles to the minimum of its potential. Yet coherent oscillations are produced at the inverse phase transition because symmetry restoration is a non-adiabatic process. 
In that case, one has
\begin{equation}
\begin{split}
M_{\chi} &\simeq 10^{-12}~\mbox{eV} \cdot B^{9/20} \left(\frac{g_* (T_{sym})}{100} \right)^{1/5} \left(\frac{g_* (T_i)}{100} \right)^{1/20} \times \\ & \times \sqrt{\frac{m_{\phi}}{10~\mbox{MeV}}}  \left(\frac{f_{p}}{30~\mbox{nHz}} \right)^{6/5}  \left(\frac{10^{-8}}{\Omega_{gw, p}} \right)^{3/20} \; ,
\label{M_x_inverse}
\end{split}
\end{equation}
where $T_{sym} \simeq m_{\phi}$ is the Universe temperature at the inverse phase transition.

We observe that in both cases GW parameters favoured by NANOGrav data imply ultra-light dark matter masses $M_{\chi}$. Notably, with these values of $M_{\chi}$, our scenario predicts superradiance instability of rotating black holes with astrophysical masses~\cite{Arvanitaki:2009fg, Brito:2015oca}. This suggests a complementary way of testing the model, in particular, the future LISA observations will probe the masses of dark matter particles corresponding to the direct phase transition, while the LIGO data may be used to test the masses involved in the inverse phase transition~\cite{Brito:2017wnc,Brito:2017zvb}.

{\it Discussions.} We have shown that the properties of GWs emitted by the network of 
melting DWs are consistent with the signal detected at PTAs. 
Keeping in mind that melting DWs do not overclose the Universe and 
the constituent field $\chi$ serves as a suitable dark matter candidate, 
this makes them interesting objects deserving further investigation. Perhaps the most important prospect for future studies is the numerical study of melting DWs evolution and eventually more precise determination 
of GW parameters, i.e., peak frequency, energy density, 
and the spectral shape including the high-frequency range. In particular, the details of the formation of melting walls and settling them into the scaling regime are yet to be better understood. The precise characteristics of those are crucial given that the most energetic GWs signals are coming from the earliest stages of the wall network evolution. With the current estimates of GW parameters, the NANOGrav signal 
is fitted in a very narrow range of model constants. Therefore, with more detailed information on the signal/improved predictions of GW properties, one will have a chance to rule out the proposed interpretation of the GW signal or establish it on firmer grounds. 
On a more theoretical side, it is interesting to embed the field $\phi$ into a realistic particle physics scenario. While in the present work we assumed that $\phi$ is in equilibrium with primordial plasma, it is worth investigating situations, where $\phi$ decouples from plasma prior to DWs formation or has never reached thermal equilibrium.

{\it Acknowledgments.} EB acknowledges support of ANR grant StronG (ANR-22-CE31-0015-01).
DG acknowledges support of the scientific program of the National  Center for Physics and Mathematics, section 5 "Particle Physics and Cosmology", stage 2023-2025. SR acknowledges the European Structural and Investment Funds and the Czech Ministry of Education, Youth and Sports (Project CoGraDS -CZ.02.1.01/0.0/0.0/15003/0000437). RS acknowledges the project MSCA-IF IV FZU - CZ.02.2.69/0.0/0.0/20 079/0017754, European Structural and Investment Fund, and the Czech Ministry of Education, Youth and Sports. AV was supported by the Czech Science Foundation (GA\v{C}R), project 20-28525S and is thankful to Enrico Barausse for discussions.

\section*{Appendix}

In the main text, to derive the low frequency behaviour of GW spectrum, i.e., $\Omega_{gw} (f) \propto f^2$, we accounted only for emission 
at the times $t$ defined from $f \sim H(t) a(t)/a_0$. Let us ensure that the contribution due to GWs emitted at earlier and later times than $t$, 
does not affect this behaviour. 
 For this purpose, we split the overall time range of GW emission in small intervals $\Delta t_k$. 
Then, the contributions of GWs at frequency $f$ emitted in these time intervals sum up to
\begin{equation}
\label{sum}
 \Omega_{gw} (f) \sim \sum_i \tilde{\Omega}_{gw, p} (\Delta t_k ) \cdot S\left(\frac{f}{\tilde{f}_{p, k}} \right) \; ,
\end{equation}
where $\tilde{\Omega}_{gw, p} (\Delta t_{k}) \equiv \tilde{\Omega}_{gw, p} (t_k+\Delta t_{k})- \tilde{\Omega}_{gw, p} (t_k)$ is the fractional energy density of GWs emitted in the interval $\Delta t_k$, and $\tilde{f}_{p, k} \sim H (t_k)a(t_k)/a_0$ is the peak frequency of these GWs; the function $S$ encoding spectral shape of GWs emitted in the interval $\Delta t_k$ will be concretized shortly. The 'tilde' notation is introduced to avoid confusion with the analogous quantities characterizing emission at the absolute peak, i.e., $f_p$ and $\Omega_{gw,p}$. For infinitely narrow intervals $\Delta t_k \rightarrow dt_k$, one replaces
\begin{equation}
\label{infinite}
\tilde{\Omega}_{gw, p} (\Delta t_k ) \rightarrow d\tilde{\Omega}_{gw, p} \; .
\end{equation}
Consequently, one casts Eq.~\eqref{sum} in the integral form: 
\begin{equation}
 \Omega_{gw} (f) \sim \int^{f_p}_{f_{min}} d \tilde{\Omega}_{gw, p} \cdot S \left(\frac{f}{\tilde{f}_p}  \right) \; ,
\end{equation}
or equivalently 
\begin{equation}
\label{int}
 \Omega_{gw} (f) \sim \frac{2\Omega_{gw,p}}{f^2_p} \int^{f_p}_{f_{min}} d \tilde{f}_p  \tilde{f}_p \cdot S \left(\frac{f}{\tilde{f}_p} \right) \; ,
\end{equation}
where we used that
\begin{equation}
\tilde{\Omega}_{gw, p} =\Omega_{gw, p} \cdot \left(\frac{\tilde{f}_p}{f_p} \right)^2 \; .
\end{equation}
Note that the lower integration limit $f_{min}$ corresponds to the moment of time, when the network of melting DWs disappears at the inverse phase transition. 
This will not play a profound role in what follows. 

For concreteness, we use the following spectral shape 
\begin{equation}
S \left(\frac{f}{\tilde{f}_p} \right) =\left(\frac{f}{\tilde{f}_p} \right)^p \cdot \frac{2}{1+ (f/\tilde{f}_p)^{p+q}} \; ,
\end{equation}
where we keep the numbers $p>0$ and $q>0$ generic for the time being. Substituting this into Eq.~\eqref{int} and switching to the integration constant $x=f/\tilde{f}_p$, we obtain
\begin{equation}
\label{pq}
\Omega_{gw} (f) \sim 4\Omega_{gw, p} \cdot \left(\frac{f}{f_p} \right)^2 \cdot \int^{f/f_{min}}_{f/f_p} \frac{x^{p-3} dx}{1+x^{p+q}} \; .
\end{equation}
For $p>2$, the integral here is convergent and it is saturated at $x \sim 1$, which means that the peaks with $\tilde{f}_p \sim f$ give the main contribution.
Being interested in the regime $f_{min}\ll f \ll f_p$, we replace
\begin{equation}
\int^{f/f_{min}}_{f/f_p} \rightarrow \int^{+\infty}_0 \; ,
\end{equation}
in which case one can take the integral explicitly:
\begin{equation}
\label{final}
\Omega_{gw} (f) \sim 4 \Omega_{gw, p} \cdot \left(\frac{f}{f_p} \right)^2 \cdot \frac{\Gamma \left(\frac{p-2}{p+q} \right) \Gamma \left(\frac{q+2} {p+q} \right)}{p+q} \; .
\end{equation}
Hence, for $p>2$ and $q>0$ we recover our $\Omega_{gw} \propto f^2$ behaviour. 

To complete the proof, recall that the exponent $p$ describes the low-frequency tail of GWs emitted during the short time interval $dt_k$ at radiation-domination. Under 
these conditions, causality arguments fix $p=3$ [35, 36, 37], which is sufficient for convergence of the integral in Eq.~\eqref{pq}. The value of $q$ is fixed by simulations with constant tension domain walls to be $q=1$; however, we do not need to assume this, as convergence of the integral~\eqref{final} is warranted for arbitrary $q>0$. Having this said, 
we confirm the low frequency behaviour $\Omega_{gw} \propto f^2$.

\end{document}